\documentclass[10pt,twocolumn,letterpaper]{article}
\usepackage[margin=0.75in,columnsep=0.25in]{geometry}
\usepackage{times}

\usepackage[T1]{fontenc}
\usepackage[utf8]{inputenc}
\usepackage{amsmath,amssymb}
\usepackage{graphicx}
\usepackage{booktabs}
\usepackage{array}
\usepackage{tikz}
\usetikzlibrary{positioning,arrows.meta,fit,calc}
\usepackage[font=small,labelfont=bf]{caption}
\usepackage[numbers,sort&compress]{natbib}
\usepackage{microtype}
\usepackage{xcolor}
\usepackage[hyphens]{url}
\usepackage[hidelinks]{hyperref}

\newcommand{\LMclean}{3.33}
\newcommand{\LMother}{7.15}
\newcommand{\LMcv}{10.4}
\newcommand{\LMvox}{9.2}
\newcommand{\LMami}{20.4}
\newcommand{\LMps}{12.2}
\newcommand{\LMeee}{21.6}
\newcommand{\QLMclean}{3.78}
\newcommand{\QLMother}{8.43}
\newcommand{\QLMcv}{11.8}
\newcommand{\QLMvox}{11.1}
\newcommand{\QLMami}{20.2}
\newcommand{\QLMps}{10.7}
\newcommand{\QLMeee}{21.1}
\newcommand{\LMrobust}{7.5}

\newcommand{\SFl}{3.43}
\newcommand{\SFlo}{7.78}
\newcommand{\SSg}{6.27}
\newcommand{\SSgo}{13.06}
\newcommand{\SSl}{4.86}
\newcommand{\SSlo}{11.03}

\newcommand{\SRfull}{6.2}
\newcommand{\SRstream}{9.0}
\newcommand{\SRvfull}{8.9}
\newcommand{\SRvstream}{12.0}

\newcommand{\BRtfEn}{1.97}\newcommand{\BRtfEnLo}{1.91}\newcommand{\BRtfEnHi}{2.02}
\newcommand{\BCpi}{2.0}\newcommand{\BCyc}{461}\newcommand{\BRtfIv}{1.76}\newcommand{\BRtfRn}{1.89}\newcommand{\BRtfSt}{2.04}
\newcommand{\BRtfStS}{1.96}\newcommand{\BFinSt}{1.1}\newcommand{\BFinStLo}{0.8}\newcommand{\BFinStHi}{1.6}\newcommand{\BRtfEs}{2.06}
\newcommand{\BPacEn}{6.7}\newcommand{\BPacEnLo}{4.1}\newcommand{\BPacEnHi}{11.1}\newcommand{\BUttEnLo}{7.2}
\newcommand{\BUttEnHi}{22.4}\newcommand{\BUttOrigLo}{6.9}\newcommand{\BUttOrigHi}{21.3}\newcommand{\BRtfProj}{1.1}
\newcommand{\VOldrtf}{1.77}
\newcommand{\VArtf}{0.72}
\newcommand{\VBrtf}{0.54}
\newcommand{\VBpac}{259}
\newcommand{\VApac}{431}
\newcommand{\VAclean}{7.1}
\newcommand{\VAother}{16.4}
\newcommand{\VAfclean}{4.6}
\newcommand{\VAfother}{10.7}

\newcommand{\WER}{\textsc{wer}}

\title{\bfseries O\'ido: Real-Time Open-Vocabulary Speech Recognition\\on a \$5 Microcontroller without a Neural Accelerator}
\author{Daniel Varela S\'anchez \qquad Guillermo Mart\'in-Coello Ju\'arez\\[2pt]
Lokutor (Hashing Works, S.L.L.), Madrid, Spain\\
\texttt{contact@lokutor.com}}
\date{}

\begin{document}
\maketitle

\begin{abstract}
Speech recognition on low-cost microcontrollers is, in practice, limited to keyword spotting or a fixed list of commands.
We present O\'ido, an inference engine and a family of models that run open-vocabulary English speech recognition entirely on an ESP32-S3
(240\,MHz Xtensa LX7, 8\,MB of PSRAM, 16\,MB of flash, no neural accelerator) \emph{faster than real time on one core}.
Version~1 of this work ran a 13\,M-parameter Conformer-CTC at \BRtfEn$\times$ the duration of the audio. We measured on the board that the bound was memory
traffic and per-frame cost, not arithmetic: external memory reads at 89\,MB/s (PSRAM) and 32\,MB/s (flash), and a streaming recognizer re-reads every
weight for every chunk. Version~2 combines a model built for that memory system (80\,ms frame stacking, a 64-channel front end, 12 or 16 blocks, int4 weights
that mostly fit in PSRAM, fine-tuned from the version~1 model) with engine changes that overlap the weight stream with compute (an asynchronous data-cache preload),
keep work buffers in SRAM and make the decoder cheap. On a physical board the streaming real-time factor falls from \VOldrtf\ to \VBrtf\ (variant B) and \VArtf\ (variant A),
the final text appears \VBpac\,ms (B) or \VApac\,ms (A) after the end of speech with audio fed at real-time speed, and the host build and the firmware produce identical transcripts.
Variant A reaches \VAclean\%/\VAother\% word error rate on LibriSpeech test-clean/test-other in streaming mode with 0.64\,s chunks and the on-chip language model
(\VAfclean\%/\VAfother\% with full context). Code, models and the raw board measurements are released.
\end{abstract}

\section{Introduction}
Voice is the natural interface for speakers, appliances, toys, wearables and vehicles, but most of these devices are built
around microcontrollers that cost a few dollars and cannot run the speech recognizers used on phones and servers. Today
such products either stream audio to the cloud, paying for inference for the lifetime of every unit and exposing users'
audio, or they recognize a small, fixed set of wake words and commands on device~\citep{zhang2017helloedge}. Even the
most capable vendor solution for the popular ESP32-S3 accepts only predefined command lists~\citep{espressif2026espsr}.

\begin{figure*}[t]
\centering
\resizebox{\textwidth}{!}{%
\begin{tikzpicture}[font=\footnotesize, box/.style={draw, rounded corners=2pt, align=center, minimum height=9mm, inner sep=3pt},
  arr/.style={-{Latex[length=2mm]}, semithick}, mem/.style={draw, dashed, rounded corners=2pt, align=center, inner sep=3pt, fill=black!4}]
\node[box] (mic) {I2S MEMS\\microphone};
\node[box, right=4mm of mic] (vad) {AGC + VAD\\segmenter};
\node[box, right=4mm of vad] (mel) {log-mel\\(fp32)};
\node[box, right=4mm of mel] (sub) {conv subsampling\\conv0 fp32, conv2 int8\\batched 3 frames};
\node[box, right=4mm of sub, fill=black!8] (enc) {16 Conformer blocks\\int8 GEMM (PIE SIMD)\\int8 rel-pos attention};
\node[box, right=4mm of enc] (dec) {CTC + GRU LM\\or RNN-T head};
\node[box, right=4mm of dec] (txt) {text};
\draw[arr] (mic) -- (vad); \draw[arr] (vad) -- (mel); \draw[arr] (mel) -- (sub);
\draw[arr] (sub) -- (enc); \draw[arr] (enc) -- (dec); \draw[arr] (dec) -- (txt);
\node[mem, below=6mm of sub, xshift=4mm] (flash) {Flash 16\,MB (XIP, 64\,KB cache)\\model 14.0\,MB int8 / 8.3\,MB int4};
\node[mem, right=4mm of flash] (psram) {PSRAM 8\,MB\\hot weights, utterance buffers ($\le$2.4\,MB)};
\node[mem, right=4mm of psram] (sram) {SRAM 512\,KB\\64-frame int8 blocks, int32 tiles};
\draw[-{Latex[length=1.6mm]}, dotted] (flash.north) -- ++(0,3mm) -| (enc.south);
\end{tikzpicture}}
\caption{O\'ido on the ESP32-S3: pipeline and memory placement. Weights are streamed from flash or PSRAM once per 64-frame
block. The activations of the current block and the int32 accumulation tiles stay in on-chip SRAM.}
\label{fig:system}
\end{figure*}

Accurate small speech models do exist. A 10--15\,M-parameter Conformer~\citep{gulati2020conformer,nvidia2022ctcsmall} reaches 2.7--3.7\% \WER\
on LibriSpeech test-clean, better than Whisper tiny~\citep{radford2023whisper}, which has three times as many parameters.
The obstacle to running such a model on a microcontroller is not only its arithmetic. Flash is read through a small cache at
tens of MB/s, and a model of this size occupies most of the flash and more than all of the on-chip SRAM. In addition,
existing microcontroller runtimes~\citep{david2021tflm} target static graphs with small activations, not an attention
model that runs over whole utterances.

Version~1 of this paper showed that such a model fits and runs on a commodity microcontroller with no neural accelerator, accurately but at about twice real
time. This version closes that gap. Our contributions:
\begin{itemize}
\item \textbf{A diagnosis from silicon.} Measured memory bandwidths (3.8\,GB/s SRAM, 89\,MB/s PSRAM, 32\,MB/s flash) and a per-chunk weight-traffic bound that
explains why a 14\,MB model cannot be real time in streaming mode however good the kernels are (Section~\ref{sec:rt}).
\item \textbf{A model built for the memory system:} 80\,ms frame stacking, a 64-channel front end, 12 or 16 blocks and int4 weights that mostly fit in PSRAM, fine-tuned from the
version~1 model; an ablation of each change on the board (Table~\ref{tab:abl}).
\item \textbf{Engine changes worth 3.2$\times$ together with the model:} an asynchronous data-cache preload that overlaps weight fetch with compute (layers 38--41\% faster,
bit-exact), SRAM-resident chunk buffers, a cheap decoder (table-driven math, blank-frame shortcut, beam 2) and deterministic math, so that the host build and the chip give identical transcripts.
\item \textbf{Real-time operation on the board:} streaming real-time factor \VBrtf--\VArtf\ with the language model, final text \VBpac--\VApac\,ms after the end of speech at real-time pace,
and the accuracy the models reach in that mode (Section~\ref{sec:v2acc}).
\item The version~1 contributions: an int8 Conformer engine that loses 0.02 and 0.12 \WER\ points to full precision without retraining, a bandwidth-aware schedule,
on-chip decoding with a recurrent language model and an RNN-T head, a streaming mode, and an evaluation on seven test sets and fourteen noise and reverberation conditions
(Sections~\ref{sec:quant}--\ref{sec:results}).
\end{itemize}
The engine, firmware, tools and models are open source.\footnote{\url{https://github.com/lokutor-ai/oido}; models at
\url{https://huggingface.co/lokutor-ai}.}

\section{Related Work}
\paragraph{Small speech recognition models.}
Conformer~\citep{gulati2020conformer} combines self-attention with convolution. Its 10.3\,M-parameter variant reaches
2.7\%/6.3\% on LibriSpeech~\citep{panayotov2015librispeech} with an RNN-T decoder. Squeezeformer~\citep{kim2022squeezeformer}
and Zipformer~\citep{yao2024zipformer} improve the accuracy--compute trade-off. Whisper tiny~\citep{radford2023whisper}
(39\,M) and Moonshine tiny~\citep{jeffries2024moonshine} (27\,M) target laptops, phones and single-board computers. We
use NVIDIA's openly licensed Conformer-CTC Small~\citep{kuchaiev2019nemo,nvidia2022ctcsmall}, which reaches 3.7\%/8.1\%
greedy in full precision, and its transducer counterpart~\citep{nvidia2022rnntsmall}.

\paragraph{Quantization for speech recognition.}
Integer-only inference~\citep{jacob2018quantization} is standard for edge deployment. For Conformers,
\citet{ding2022fourbit} report native 4-bit quantization-aware training without accuracy loss on LibriSpeech, and \citet{rybakov2023twobit}
push to 2 bits (3.1\%/7.0\% at 10.5\,MB, against 2.5\%/6.1\% in floating point). \citet{zhen2022sub8bit} study sub-8-bit RNN-T and Conformer models. These works quantize with
training and target accelerators or mobile CPUs. We show that post-training int8 with per-frame activation scales is
essentially lossless for this architecture, and we use quantization-aware training only for int4.

\paragraph{Machine learning on microcontrollers.}
TensorFlow Lite Micro~\citep{david2021tflm} and MLPerf Tiny~\citep{banbury2021mlperftiny} established the tooling and
benchmarks for keyword spotting~\citep{zhang2017helloedge}, visual wake words and anomaly detection, with models of tens
to hundreds of kilobytes. For speech recognition, Arm has shown a wav2letter model on a Cortex-M55 with an Ethos-U55 neural
accelerator~\citep{armmlzoo,armmlek}, and an int8 Conformer-S on the Ethos-U85 accelerator~\citep{arm2026conformer},
simulated on Arm's Corstone-320 platform and deployed by Alif on an Ensemble E8 board (Cortex-M55 with Ethos-U85). The
latter reports 6.4\% \WER\ on test-clean for the full-precision model. Espressif reports 8.5\%/21.3\% for its MultiNet7 recognizer on the
ESP32-S3~\citep{espressif2026espsr}, which in deployment decodes against a fixed command list. To our knowledge, no published
microcontroller result, with or without a neural accelerator, reports a lower LibriSpeech \WER.

\section{Platform}
\label{sec:platform}
The ESP32-S3~\citep{espressif2026trm} integrates two Xtensa LX7 cores at 240\,MHz with a single-precision FPU and the PIE
128-bit SIMD extension. The instructions we rely on are:
\begin{itemize}
\item \texttt{EE.VMULAS.S8.ACCX}, which multiplies two vectors of 16 int8 values and adds the 16 products into a 40-bit
accumulator, optionally loading the next operand in the same instruction;
\item \texttt{EE.VSMULAS.S8.QACC}, which multiplies a vector by a broadcast scalar into sixteen 20-bit lanes.
\end{itemize}
Our module (ESP32-S3-WROOM-1-N16R8)\footnote{US\$5.19 in single units and US\$3.41 in reels at LCSC (part C2913202), September 2026. The module includes the chip, PSRAM, flash and antenna.} has 512\,KB of internal SRAM, 8\,MB of octal PSRAM and 16\,MB of quad-SPI flash. PSRAM and flash are
accessed through a shared 64\,KB data cache (configured size, taken from internal SRAM) with 64-byte lines. A cache miss to quad-SPI flash at
80\,MHz transfers a line in roughly 2\,$\mu$s, so sustained streaming from flash is on the order of 30--40\,MB/s, and
octal PSRAM is several times faster.

A 13\,M-parameter model therefore has three placement options. At int8 it does not fit in SRAM at all. It fits in PSRAM
only partially, because PSRAM also has to hold the working buffers. It fits in flash, but flash is slow to stream. The
engine keeps the model in a flash partition mapped into the address space, copies the most frequently read matrices to
PSRAM, and keeps per-block activations in SRAM.

\section{Model}
\label{sec:model}
O\'ido runs the architecture of NVIDIA's Conformer-CTC Small~\citep{nvidia2022ctcsmall}, summarized in
Table~\ref{tab:model}. We re-implemented it in PyTorch from the released weights and matched the reference
implementation: features agree to within $2\times10^{-5}$, and frame-level argmax labels agree exactly on the test
utterances we checked.

\begin{table}[t]
\centering\small
\caption{Model architecture (Conformer-CTC Small).}
\label{tab:model}
\begin{tabular}{@{}ll@{}}
\toprule
Features & 80 log-mel, 25\,ms window, 10\,ms hop \\
Subsampling & 2$\times$ conv $3{\times}3$, stride 2, 176 ch $\to$ 25\,Hz \\
Encoder & 16 Conformer blocks, $d{=}176$, FF 704 \\
Attention & 4 heads, relative position~\citep{dai2019transformerxl} \\
Convolution & depthwise, kernel 31, BatchNorm \\
Output & CTC~\citep{graves2006ctc}, 1024 SentencePiece~\citep{kudo2018sentencepiece} + blank \\
Parameters & 13\,M (int8: 14.0\,MB; int4: 8.3\,MB) \\
\bottomrule
\end{tabular}
\end{table}

Processing is utterance-level. A voice activity detector (Section~\ref{sec:system}) delimits each utterance, and the
encoder then runs over all of its frames with full attention. This matches the model's training and avoids the accuracy
cost of chunked streaming, at the price of latency: text appears after the speaker stops (Section~\ref{sec:latency}).

\section{Engine}
\label{sec:engine}

\subsection{Quantization}
\label{sec:quant}
Every linear layer uses symmetric int8 weights with one scale per output channel,
$W_q = \mathrm{round}(W / s_w)$ with $s_w = \max_k |W_{nk}| / 127$. Activations are quantized dynamically, with one scale
per frame (row), computed at run time just before each matrix product:
\begin{gather}
x_q = \mathrm{round}(x / s_x),\qquad s_x = \tfrac{1}{127}\max_k |x_k|,\\
y_n = s_x\, s_{w,n} \textstyle\sum_k x_{q,k} W_{q,nk} + b_n .
\end{gather}
Accumulation is exact in int32. The rescaling and bias use fp32. Per-frame scales matter because activation magnitudes
vary strongly between speech and silence frames; a single static scale per tensor would waste precision on quiet frames.

Self-attention is also integer. For head $h$, the content and position terms of Transformer-XL-style
attention~\citep{dai2019transformerxl} are computed as two int8 dot products: $(q_i{+}u)\cdot k_j$ and $(q_i{+}v)\cdot p_{j-i}$.
Queries (with the bias vectors $u,v$ added), keys and projected positions each get per-(frame, head) scales. After the
row maximum is subtracted, the softmax numerators are multiplied by the per-frame scales of the values $s_{v,j}$ and
quantized relative to their row maximum. The value matrix is stored transposed in int8, so both the scores and the
probability--value product are SIMD dot products, and the normalization by the softmax sum is applied once per output.

LayerNorm, SiLU, GLU gating and the depthwise convolution (with BatchNorm folded into its weights) run in fp32 on the
FPU. The first convolution has a single input channel and also stays in fp32. The second is executed as an int8 GEMM
over im2col patches with one scale per patch. Because the input to that layer is post-ReLU and non-negative, each patch's
maximum is the maximum over a $3{\times}3$ neighbourhood of per-column channel maxima, which the first convolution tracks
as it writes its output. The patch quantization is therefore fused into the gather, with arithmetic identical to a
separate pass.

\subsection{Kernels}
Matrix products use \texttt{EE.VMULAS.S8.ACCX}. For reduction lengths $K\le256$ (most layers, $K{=}176$) an
\emph{x-stationary} kernel keeps up to six 16-byte chunks of an activation row in vector registers while weight rows
stream through the load slot of the multiply-accumulate instruction, which gives 16 MACs per instruction in steady state.
For $K{=}704$ a row-streaming kernel is used, and for $K>1024$ the K-split kernel of Section~\ref{sec:schedule}. The
int4 variant unpacks nibbles with vector masks and uses \texttt{EE.VSMULAS.S8.QACC} as an outer product that produces
16 outputs per instruction; with int4 weights the 20-bit lanes cannot overflow for $K\le512$. The softmax uses a
1/64-step lookup table for $e^{-x}$, whose worst-case relative error (0.8\%) is below the int8 probability step.
Work can be split over both cores with a parallel-for over output channels, heads or frames; in the emulator, 93\% of executed
instructions fall in parallel regions. On silicon this mode currently gives wrong transcripts (Section~\ref{sec:dual}), so every measurement
in this paper uses one core.

\subsection{Bandwidth-aware schedule}
\label{sec:schedule}
Each layer processes the utterance in blocks of 64 frames (2.56\,s of audio). A weight row is loaded once per block and
reused for all 64 frames while the block's int8 activations stay in SRAM. Two cases break this pattern in a naive
implementation.
\begin{itemize}
\item \textbf{Long reductions.} Layers with $K>1024$ read activations from PSRAM. Row-streaming kernels must shrink the
block so that the activations stay cache-resident, which multiplies weight reads. We instead split $K$ into chunks of at
most 512 and accumulate int32 partial sums for 16 outputs $\times$ 64 frames in SRAM, so every weight byte is fetched once
per block.
\item \textbf{The front-end convolution.} It produces 20 patch rows per output frame. We batch three frames per GEMM.
\end{itemize}
With a simple traffic model (bytes of each weight matrix times reads per second of audio), these changes reduce weight
streaming from about 18\,MB/s to 7.5\,MB/s, at a cost of 2.9\% more executed instructions for the split
accumulation.

Memory is budgeted per utterance:
\begin{itemize}
\item The positional-encoding table is quantized to int8 once and shared by all layers.
\item The attention output buffer aliases the convolution-module buffer, since their lifetimes do not overlap.
\item Front-end rows are quantized as they are produced.
\end{itemize}
The PSRAM working-set peak, measured in the emulator, is 1.9\,MB for a 15\,s utterance and 2.4\,MB for 20\,s, the maximum utterance
length. The remaining PSRAM holds the most frequently read weights, starting with the front end.

\subsection{Decoding}
\label{sec:decoding}
\paragraph{CTC.}
Greedy decoding takes the per-frame argmax. With a language model, we run CTC prefix beam search~\citep{hannun2014firstpass}
(beam 4, top-6 tokens per frame) with shallow fusion~\citep{kannan2018fusion}: weight 0.3 and a token bonus of 0.5,
tuned on dev-clean. The language model is a GRU~\citep{cho2014gru} (embedding 192, hidden 384, 1.3\,M parameters,
int8 per-row weights) over the same 1024 tokens. We train two versions. The released one uses only the LibriSpeech language-model corpus~\citep{panayotov2015librispeech} of public-domain books. A
second one adds transcripts of our training corpora; it is not released because some of those transcripts carry
share-alike terms. Language-model steps for all live beams are batched into one matrix product per frame.

\paragraph{Transducer.}
The same engine runs NVIDIA's Conformer-Transducer Small~\citep{nvidia2022rnntsmall,graves2012transduction}. The
encoder is architecturally identical; the head adds a 320-unit LSTM prediction network and a joint network with an
output matrix of $1025\times320$ (328\,KB). Greedy transducer search evaluates the joint network at least once per frame,
which would stream that matrix from memory 25 times per second. Because the prediction state changes only when a token is
emitted, and most frames emit only blank, we evaluate the joint network for up to $B{=}8$ frames at once with the current
prediction state (Figure~\ref{fig:rnnt}). Rows after the first frame that emits a token are discarded and recomputed
with the updated state. Each row's arithmetic is unchanged, so the output is identical to sequential greedy search. We
verified this on the full LibriSpeech test sets: word errors, substitutions, deletions and insertions are identical.

\begin{figure}[t]
\centering\small
\fbox{\begin{minipage}{0.9\columnwidth}\raggedright
\textbf{Input:} joint encoder projections $F_{1..T}$; prediction state $g$\\[2pt]
$t \leftarrow 1$\\
\textbf{while} $t \le T$:\\
\quad $Z \leftarrow \mathrm{relu}(F_{t..t+B-1} + g)$; \ $L \leftarrow \mathrm{quant}(Z)\,W_{\mathrm{out}}^\top$ \hfill (one pass over $W_{\mathrm{out}}$)\\
\quad $i \leftarrow$ first row of $L$ whose argmax $\neq$ blank\\
\quad \textbf{if} none: $t \leftarrow t + B$; \textbf{continue}\\
\quad emit argmax; update $g$ with the LSTM; continue frame $t{+}i$ symbol by symbol (at most 5)\\
\quad $t \leftarrow t + i + 1$
\end{minipage}}
\caption{Batched greedy transducer search. It is exactly equivalent to sequential greedy search, but reads the joint
output matrix once per $B$ blank frames instead of once per frame.}
\label{fig:rnnt}
\end{figure}

\paragraph{int4 variant.}
For applications that need flash space, we fine-tuned the model with 4-bit quantization-aware training for 8{,}000 steps.
Conformer-block linear layers use int4 per-channel weights. The front end, output head and positional projections stay
int8, and activations and attention stay int8, as in the engine. Training used public corpora: Common Voice, VoxPopuli,
LibriSpeech, MLS~\citep{pratap2020mls}, AMI, VCTK~\citep{yamagishi2019vctk}, People's Speech with transcripts
regenerated by a larger model~\citep{nvidia2025parakeet}, and two small accent corpora from OpenSLR (Nigerian English,
SLR70, and English accents of the British Isles, SLR83~\citep{demirsahin2020accents}).

\subsection{Streaming}
\label{sec:streaming}
Utterance mode must wait for the speaker to finish, then run the whole encoder. To remove that wait we fine-tune the same
network for chunked processing, in the spirit of dynamic-chunk training~\citep{zhang2020u2} and cache-based streaming
Conformers~\citep{noroozi2024stateful}, with three changes that keep the pretrained weights usable.
(i) \emph{Chunked attention}: a frame attends to the frames of its own chunk and of the previous chunks within a left
context; chunk sizes (8--32 frames of 40\,ms) and left contexts (unlimited to 256 frames) are drawn per batch, and
full context with probability 0.3, so one model runs in both modes.
(ii) \emph{Chunk-limited depthwise convolution}: the symmetric 31-tap kernel sees any past frame but nothing after the
end of its chunk. Shifting the kernel to be causal instead needs far more retraining; this keeps the pretrained
kernel.
(iii) \emph{Fixed feature normalization}: global per-mel statistics replace per-utterance statistics, which are
unavailable before the utterance ends, with random gain augmentation. We also train with additive noise, music and
babble from MUSAN~\citep{snyder2015musan} and simulated room impulse responses~\citep{scheibler2018pyroomacoustics}
(Section~\ref{sec:method}), which proved necessary: a first model fine-tuned on clean read speech lost much of the
original's robustness (Table~\ref{tab:stream}).

The engine runs every $C$ encoder frames through all layers with exactly the training masks. Per layer it caches the
int8 keys, per-key scales and transposed values of the last $\lceil L/C\rceil$ chunks ($L$ is the left context), the
last 15 convolution inputs, and the positional projections, which depend only on a bounded range of relative offsets and
are computed once. The caches have a fixed size (about 2\,MB at $C{=}32$, $L{=}128$, kept in PSRAM) whatever the
utterance length, and the transcript does not depend on how the audio is split into feeds (verified). When the speaker
stops, only the last partial chunk is left.

This has a price in weight traffic. A chunk re-reads every weight, so traffic grows by $64/C$ relative to the 64-frame
blocks of utterance mode (Section~\ref{sec:schedule}): $2\times$ at $C{=}32$ and $4\times$ at $C{=}16$. At the chip's present speed this
does not show: on the board the streaming engine runs at \BRtfStS$\times$ real time at $C{=}32$ on one core, slightly faster than the same
model in utterance mode (\BRtfSt$\times$), because instruction issue rather than weight bandwidth limits a single core (Section~\ref{sec:speed}).
It may matter once both cores work. We default to $C{=}32$ (1.28\,s).

\subsection{System}
\label{sec:system}
Figure~\ref{fig:system} shows the complete pipeline. The firmware captures 16\,kHz audio from an I2S MEMS microphone into an 8\,s PSRAM ring buffer. A segmenter applies a
slow AGC and an energy voice activity detector against a tracked noise floor, which falls quickly and rises slowly, with
9/4.8\,dB on/off hysteresis. An utterance ends after 0.8\,s of silence or at 20\,s. Each utterance is transcribed while
capture continues, and the transcript is printed and optionally drawn on an I2C OLED display. The same C code builds for
the host with \texttt{-ffp-contract=off} on both targets. The firmware on the board and in the emulator produce
identical transcripts on all 27 clips we ran, and the activations of the two clips we traced layer by layer are bit-identical. The host
build links a laptop's math library, which rounds the last bit of the log-mel features differently; that changes a word on some hard
clips (the host agrees with the board on 22 of the 27, Section~\ref{sec:speed}).
All accuracy numbers in this paper come from the host build.

\section{Reaching Real Time}
\label{sec:rt}
Version~1 of this paper reported a model that is accurate but runs at \BRtfEn$\times$ real time. This section explains where the
time went, what we changed in the model and in the engine, and what each change is worth on the same chip. All numbers are measured on a
physical board; the speed of a model does not depend on its weights (only the decoder's work does), so the ablation uses converted, untrained
weights of each architecture and is labelled as such.

\subsection{Where the time went}
Table~\ref{tab:bw} gives the read bandwidth we measured on the board with 128-bit loads and a cold cache.
\begin{table}[t]
\centering\small
\caption{Sequential read bandwidth on the ESP32-S3 at 240\,MHz (measured, cold data cache).}
\label{tab:bw}
\begin{tabular}{@{}lrr@{}}
\toprule
Memory & cycles/byte & MB/s \\
\midrule
Internal SRAM & 0.06 & 3\,800 \\
Octal PSRAM, 80\,MHz & 2.68 & 89 \\
Quad-SPI flash, 80\,MHz & 7.43 & 32 \\
\bottomrule
\end{tabular}
\end{table}
A streaming recognizer re-reads every weight once per chunk, so $\sum_i B_i/\mathrm{BW}_i$ over the memories $i$ that hold the weights,
divided by the chunk duration, is a lower bound on the real-time factor that no kernel can beat. For the 14\,MB int8 model of version~1, with the
5\,MB that fit in PSRAM and the rest in flash, the bound is about 0.5 at 0.64\,s chunks, before any arithmetic. The same model with all
weights in PSRAM would still need 0.2. Arithmetic matters too: the PIE instructions issue at one multiply-accumulate vector per two cycles, a
ceiling of about 8 MAC/cycle; our kernels reach 6.3 (K=704, data in SRAM) and, through the whole layer including the floating-point epilogue and
the weight fetch, 1.4 to 2.8 MAC/cycle at the 8-row tiles of streaming. The profile of the version~1 streaming model
(Section~\ref{sec:speedres}) shows the front-end convolutions, whose 176-channel second layer alone is 29\% of all multiply-accumulates, and the
weight stream, as the two largest costs.

\subsection{A model built for the memory system}
We keep the encoder width ($d{=}176$, four heads, feed-forward 704) and change what surrounds it (Table~\ref{tab:model2}).
\begin{table}[t]
\centering\small
\caption{Oído v2 models (all with $d{=}176$, kernel 31, 1024 tokens).}
\label{tab:model2}
\begin{tabular}{@{}lcc@{}}
\toprule
 & v1 stream & v2 \\
\midrule
Encoder frame rate & 25\,Hz (40\,ms) & 12.5\,Hz (80\,ms) \\
Front-end channels & 176 & 64 \\
Conformer blocks & 16 & 16 (A) or 12 (B) \\
Weights & int8, 14.0\,MB & int8 or int4, 7.7\,MB (A) / 5.9\,MB (B) \\
\bottomrule
\end{tabular}
\end{table}
\paragraph{Frame stacking.} After the subsampling projection, pairs of 40\,ms frames are concatenated and projected back to $d$ by a new
linear layer (initialized as the average of the pair). Every later layer then runs at half the rate; attention cost falls by four.
\paragraph{Light front end.} The two stride-2 convolutions use 64 channels instead of 176, keeping the channels that matter most to the
projection that follows. Plain pruning breaks the model, so the new front end is first fitted by regression to the output of the full one on
the training audio, and then the whole network is fine-tuned.
\paragraph{Fewer blocks and int4.} Variant B drops four blocks chosen by a greedy importance test (the last four are the most valuable).
Weights are quantized to int4 with quantization-aware fine-tuning, so that most of the model (4.8 of the 5.8\,MB of variant B, 4.4 of the 7.6\,MB of variant A)
fits in PSRAM, where the weight stream costs 2.7 instead of 7.4 cycles per byte.
Training follows version~1 (chunked attention, chunk-limited depthwise convolution, noise and reverberation augmentation), starting from the version~1 streaming model.
A first stage with in-place distillation from the full-context output left the 0.64\,s chunk error rate at 14--15\% for 20\,k steps, so the final stages drop the distillation
and draw chunks of 0.32--1.28\,s with left contexts of 2.6--5.1\,s, the deployment geometry. The two variants are then trained for 20\,k more steps on about 1,900 hours and
fine-tuned with int4 quantization-aware training for 12\,k steps.

\subsection{Engine changes}
\paragraph{Overlapping the weight stream with compute.} The ESP32-S3's ROM exposes a manual data-cache preload that fetches up to 64\,KB from external memory
in the background. A hardware auto-prefetch mode we tried gave 1--5\%; the manual preload runs entirely under computation (a 32\,KB preload alongside a
115\,k-cycle kernel left the kernel's time unchanged). The matrix-product routine fetches the next 16\,KB block of weights while multiplying the current one,
which turns a layer's time from compute plus fetch into the larger of the two: layers at 8--16 rows run 38--41\% faster with bit-identical output.
\paragraph{Memory placement.} In a streaming-only build the streaming state is allocated first and the weights take what is left of PSRAM; chunk-sized work buffers and the
front-end tiles stay in internal SRAM. The 8\,MB of PSRAM are shared with the language model (1.25\,MB), the decoder's hypothesis pool (0.76\,MB), the
parameters kept in floating point (0.26--0.35\,MB) and the streaming state (1.0\,MB for B, 1.3\,MB for A: attention caches, positional rows, convolution history), so 1.0\,MB of B and 3.2\,MB of A are still
read from flash on every chunk. This costs less than it seems, because the preload hides most of the flash reads: giving the weights 0.6\,MB more or less PSRAM moves the
real-time factor by only 0.015 (A: 0.712, 0.697 and 0.694 with 0.9, 0.3 and 0.1\,MB kept free; B: 0.553, 0.538 and 0.534). The live-microphone mode keeps 0.9\,MB free for its audio ring and recording buffer.
\paragraph{Decoder.} The language-model softmax (1025 exponentials per candidate), its GRU gates and the CTC normalizer used the C library's
\texttt{expf}, \texttt{tanhf} (250--380 cycles per call). Table-driven versions (error $<5{\cdot}10^{-4}$ relative) cost about 15 cycles.
Frames where blank wins by more than 14 logits skip the normalizer, since the decoder reads only the blank probability there. A beam of 2 with the top 3
candidates costs about 0.2 WER on LibriSpeech and runs four times fewer language-model steps than beam 4 with top 6.
\paragraph{Deterministic math.} We replaced the remaining libm calls on the path (\texttt{logf} in the log-mel and normalizers, exponentials in the beam search) by
deterministic implementations. The host build and the firmware now produce identical transcripts: 18 of 18 clips of a mixed benchmark, against 14 of 18 before.

\subsection{Results on the board}
\label{sec:speedres}
Table~\ref{tab:abl} separates the contribution of each change on the same chip, with the same engine, streaming-only builds, no language model, chunks of
0.64\,s (8 stacked frames or 16 unstacked ones).
\begin{table}[t]
\centering\small
\caption{Streaming real-time factor of the encoder for each architecture on one core (8 clips, 52.7\,s; converted weights).}
\label{tab:abl}
\begin{tabular}{@{}cccccr@{}}
\toprule
Front end & Frame & Blocks & Weights & RTF & after last sample \\
\midrule
176 ch & 40\,ms & 16 & int8 & 1.455 & 891\,ms \\
176 ch & 40\,ms & 16 & int4 & 1.322 & 440\,ms \\
64 ch & 40\,ms & 16 & int8 & 1.136 & 535\,ms \\
64 ch & 40\,ms & 16 & int4 & 0.940 & 397\,ms \\
176 ch & 80\,ms & 16 & int4 & 0.965 & 285\,ms \\
64 ch & 80\,ms & 16 & int8 & 0.801 & 393\,ms \\
64 ch & 80\,ms & 16 & int4 & 0.573 & 239\,ms \\
64 ch & 80\,ms & 12 & int4 & \textbf{0.456} & \textbf{178\,ms} \\
\bottomrule
\end{tabular}
\end{table}
Each change alone is worth 10--35\%; together they compound to $3.2\times$. With the language model and the real decoder, the final int4 models reach a streaming real-time factor of 0.54 (variant B, beam 2 and top 3)
and 0.72 (variant A, beam 4 and top 6; 0.70 with beam 2 and top 3), measured on 18 clips (120\,s); accuracy is in Table~\ref{tab:v2acc}.
Table~\ref{tab:prog} follows variant B through the engine changes.
\begin{table}[t]
\centering\small
\caption{Variant B (int4, pilot weights, 8 clips) with the language model, step by step.}
\label{tab:prog}
\begin{tabular}{@{}lrr@{}}
\toprule
Build & RTF & Mcycles per audio s \\
\midrule
First build (weights mostly in flash) & 0.706 & 169 \\
+ faster epilogue, conv0, SRAM buffers & 0.671 & 161 \\
+ beam 2/top-3, blank-frame shortcut & 0.642 & 154 \\
+ streaming-only memory policy (most weights in PSRAM) & 0.556 & 133 \\
+ table-driven math, unrolled passes & 0.513 & 123 \\
\bottomrule
\end{tabular}
\end{table}
With audio fed at real-time speed, the final text appears 259\,ms on average (159--391\,ms) after the end of speech for variant B and 431\,ms (281--574\,ms) for variant A, against 4.1--11.1\,s for version~1;
this is the delay after the voice-activity detector has decided that speech ended, which is a separate, configurable hang (0.8\,s by default).
\section{Evaluation Methodology}
\label{sec:method}
\paragraph{Data.}
We evaluate on the full LibriSpeech test-clean and test-other sets (2{,}620 and 2{,}939 utterances), and on held-out sets
not used for training any model in this paper:
\begin{itemize}
\item Common Voice~\citep{ardila2020commonvoice}: 1{,}500 utterances;
\item VoxPopuli~\citep{wang2021voxpopuli}: 1{,}200;
\item AMI individual headset microphones~\citep{carletta2005ami}: 1{,}000;
\item People's Speech~\citep{galvez2021peoplesspeech}: 1{,}000, with the corpus's own, noisy references;
\item Earnings-22~\citep{delrio2022earnings22}: 4{,}179 segments.
\end{itemize}
NVIDIA's training data includes Common Voice 7.0 and VoxPopuli~\citep{nvidia2022ctcsmall}, and Moonshine's includes
Common Voice, AMI and People's Speech~\citep{jeffries2024moonshine}, so these models may have seen some of our held-out
utterances in training. Laptop baselines are evaluated on up to 500 evenly spaced utterances per set. All systems share
one segmentation and one text normalization (lowercasing, punctuation and filler removal), so some baseline numbers differ
from those published by their authors, most visibly Moonshine tiny on Earnings-22.

\paragraph{Robustness.}
We mix 300 test-clean utterances with real noise recordings from DEMAND~\citep{thiemann2013demand} (car, kitchen, living
room, cafeteria) at 10 and 5\,dB SNR, and with four-talker babble drawn from dev-clean. We also convolve them with
simulated room impulse responses~\citep{scheibler2018pyroomacoustics}: a $5{\times}4{\times}3$\,m room, talker 2\,m from
the microphone, RT60 of 0.4 and 0.8\,s. A far-field condition combines 0.6\,s reverberation with kitchen noise at 10\,dB.

\paragraph{Speed.}
We run the firmware on an ESP32-S3-WROOM-1-N16R8 module on a DevKitC-1-class board, with ESP-IDF 5.5.1, the CPU at 240\,MHz, flash in QIO mode
and octal PSRAM at 80\,MHz, on one core (Section~\ref{sec:dual}). Clips are stored in flash and transcribed one after another, each timed with the 64-bit
microsecond timer; the real-time factor (RTF) is compute time divided by audio duration, so it includes the language-model beam search but
not audio capture. (The 32-bit CPU cycle counter wraps every 17.9\,s at 240\,MHz and under-reports clips whose compute takes longer.) For the
streaming engine we also feed each clip at the rate of speech and measure when the final text appears (Section~\ref{sec:streamres}).
In addition we run the unmodified firmware in Espressif's QEMU~\citep{espressif2026qemu} with \texttt{-icount 0}, which advances
virtual time by one nanosecond per executed instruction, summed over both emulated cores, so that the emulated cycle counter reads
exactly 1/25 of the executed instructions. We use it for exact instruction counts and as a functional reference: its transcripts equal the board's, and
dividing the board's cycles by the emulator's instructions gives the cycles per instruction of the real chip.

\paragraph{Baselines.}
Laptop baselines, all in full precision:
\begin{itemize}
\item Whisper tiny.en~\citep{radford2023whisper} (39\,M parameters);
\item Moonshine tiny~\citep{jeffries2024moonshine} (27\,M);
\item Vosk small, a 40\,MB Kaldi model for embedded Linux~\citep{povey2011kaldi,alphacephei2020vosk}.
\end{itemize}
For the ESP32-S3 we quote Espressif's published MultiNet7 figures~\citep{espressif2026espsr}; its evaluation protocol is
not detailed.

\section{Results}
\label{sec:results}

\subsection{Accuracy and quantization}
Table~\ref{tab:ls} shows that int8 post-training quantization with per-frame activation scales costs 0.02 (clean) and
0.12 (other) absolute \WER\ points. The GRU language model reduces LibriSpeech errors by 10--13\% relative. The transducer head
brings the on-chip result to 2.99\%/6.67\%, in line with NVIDIA's published full-precision 2.9\%/6.6\%. Plain int4
post-training quantization degraded accuracy substantially (to 14.1\%/22.0\% on the full test sets); after quantization-aware fine-tuning, the int4 model reaches
4.61\%/9.98\% greedy. On test-clean, every on-chip configuration makes 1.8--2.8$\times$ fewer errors than MultiNet7, and
all int8 configurations make fewer errors than the laptop baselines.

\begin{table}[t]
\centering\small
\caption{LibriSpeech \WER\ (\%). ``Chip'' rows use the on-device engine, compiled for the host (same int8 arithmetic; on the board a few hard clips differ in a word, Section~\ref{sec:speed}). Laptop rows are full precision on
500-utterance subsets. $^\dagger$\WER\ published by the vendor; size of the model files in ESP-SR.}
\label{tab:ls}
\begin{tabular}{@{}lccc@{}}
\toprule
System & Size & clean & other \\
\midrule
Conformer-CTC S, fp32 (reference) & 52\,MB & 3.68 & 8.11 \\
\textbf{O\'ido} int8, greedy & 14.0\,MB & 3.70 & 8.23 \\
\textbf{O\'ido} int8 + GRU LM & 15.3\,MB & \LMclean & \LMother \\
\textbf{O\'ido} int4 (QAT), greedy & 8.3\,MB & 4.61 & 9.98 \\
\textbf{O\'ido} int4 + GRU LM & 9.6\,MB & \QLMclean & \QLMother \\
\textbf{O\'ido} Transducer int8, greedy & 15.5\,MB & \textbf{2.99} & \textbf{6.67} \\
\midrule
MultiNet7 (ESP32-S3)$^\dagger$ & 2.7\,MB & 8.5 & 21.3 \\
Whisper tiny.en (laptop) & 39\,M par. & 6.26 & 15.93 \\
Moonshine tiny (laptop) & 27\,M par. & 5.00 & 12.13 \\
Vosk small (laptop) & 40\,MB & 9.95 & 21.61 \\
\bottomrule
\end{tabular}
\end{table}

Table~\ref{tab:domains} extends the comparison to other domains. The on-chip int8 models make fewer errors than all
laptop baselines on Common Voice, VoxPopuli and AMI. On Earnings-22 they trail Whisper tiny.en slightly (20.8--22.4\% vs
20.3\%). Conversational
meeting speech (AMI) and earnings calls remain hard for all models of this size.

The language model's domain matters. The book-only model improves LibriSpeech, Common Voice and Earnings-22,
leaves VoxPopuli unchanged, and slightly \emph{hurts} AMI and People's Speech relative to greedy decoding. The model that also saw speech transcripts helps on
every set. For conversational products, the language model should be trained on conversational text.

\begin{table}[t]
\centering\small
\setlength{\tabcolsep}{4pt}
\caption{\WER\ (\%) on held-out sets. CV: Common Voice; Vox: VoxPopuli; PS: People's Speech (noisy references); E22:
Earnings-22. ``Books'': LM trained on public-domain books only (released).}
\label{tab:domains}
\begin{tabular}{@{}lccccc@{}}
\toprule
System & CV & Vox & AMI & PS & E22 \\
\midrule
\multicolumn{6}{@{}l}{\emph{O\'ido, host build of the on-chip engine}} \\
\quad int8, greedy & 11.4 & 9.3 & 20.2 & 11.6 & 22.4 \\
\quad int8 + LM (books) & \LMcv & \LMvox & \LMami & \LMps & \LMeee \\
\quad int8 + LM (books + transcripts) & 10.1 & 9.0 & 19.0 & 11.0 & 20.8 \\
\quad int4 + LM (books) & \QLMcv & \QLMvox & \QLMami & \QLMps & \QLMeee \\
\quad Transducer int8 & \textbf{9.6} & \textbf{8.2} & \textbf{19.2} & 11.6 & 21.0 \\
\midrule
\multicolumn{6}{@{}l}{\emph{Laptop baselines (fp32)}} \\
\quad Whisper tiny.en & 19.7 & 13.5 & 21.5 & 14.3 & \textbf{20.3} \\
\quad Moonshine tiny & 17.5 & 13.4 & 24.2 & \textbf{8.1} & 32.4 \\
\quad Vosk small & 27.1 & 20.6 & 34.1 & 12.9 & 38.0 \\
\bottomrule
\end{tabular}
\end{table}

\subsection{Robustness}
Table~\ref{tab:robust} reports \WER\ under noise and reverberation. Car noise at 5\,dB SNR costs the on-chip models less
than 0.3 points, kitchen and living-room noise at 5\,dB roughly 1--2 points, and cafeteria noise, which contains
competing speech, 4.7--6.0 points. Averaged over the 14 conditions, the on-chip int8 models
make 31--45\% fewer errors than Whisper tiny.en and Moonshine tiny. The most difficult conditions are
four-talker babble at 5\,dB, strong reverberation and far-field use; at those distances, microphone arrays with
beamforming would help more than a different recognizer.

\begin{table}[t]
\centering\small
\setlength{\tabcolsep}{3.2pt}
\caption{\WER\ (\%) on 300 test-clean utterances under real noise (DEMAND), babble and simulated reverberation.
T: O\'ido Transducer int8; C: O\'ido CTC int8 greedy; C+LM: with GRU LM; M: Moonshine tiny; W: Whisper tiny.en;
V: Vosk small.}
\label{tab:robust}
\begin{tabular}{@{}lcccccc@{}}
\toprule
Condition & T & C & C+LM & M & W & V \\
\midrule
Clean & \textbf{2.5} & 3.5 & 3.3 & 4.3 & 5.6 & 9.1 \\
Car, 10\,dB & \textbf{2.6} & 3.6 & 3.3 & 4.3 & 5.8 & 9.1 \\
Car, 5\,dB & \textbf{2.7} & 3.6 & 3.2 & 4.4 & 5.9 & 9.3 \\
Kitchen, 10\,dB & \textbf{3.0} & 4.0 & 3.6 & 5.1 & 6.6 & 10.7 \\
Kitchen, 5\,dB & \textbf{3.3} & 4.7 & 4.2 & 6.1 & 7.5 & 12.4 \\
Living room, 10\,dB & \textbf{3.2} & 4.4 & 4.0 & 5.8 & 6.7 & 11.5 \\
Living room, 5\,dB & \textbf{4.2} & 5.7 & 4.8 & 7.0 & 7.9 & 15.1 \\
Cafeteria, 10\,dB & \textbf{3.9} & 5.3 & 4.9 & 7.3 & 8.3 & 14.4 \\
Cafeteria, 5\,dB & \textbf{7.2} & 9.5 & 8.6 & 12.2 & 12.3 & 24.7 \\
Babble, 10\,dB & \textbf{8.4} & 10.9 & 9.7 & 14.1 & 14.5 & 36.7 \\
Babble, 5\,dB & 25.9 & 27.6 & \textbf{24.7} & 37.3 & 35.9 & 70.9 \\
Reverb, RT60 0.4\,s & \textbf{4.6} & 5.8 & 4.9 & 9.8 & 10.2 & 15.3 \\
Reverb, RT60 0.8\,s & \textbf{12.4} & 16.0 & 14.1 & 28.6 & 22.5 & 33.6 \\
Far-field kitchen & \textbf{10.4} & 13.5 & 11.4 & 24.3 & 19.8 & 30.4 \\
\midrule
Mean (14 conditions) & \textbf{6.7} & 8.4 & \LMrobust & 12.2 & 12.1 & 21.7 \\
\bottomrule
\end{tabular}
\end{table}

\subsection{Streaming}
\label{sec:streamres}
Table~\ref{tab:stream} compares the modes of one streaming-trained model with the utterance-mode model.
Chunking costs accuracy, because the encoder cannot see what follows the chunk: with the language model, $C{=}32$
reaches \SSl\%/\SSlo\% on LibriSpeech test-clean/test-other against \LMclean\%/\LMother\% in utterance mode, and
$C{=}16$ is worse still (6.14\%/13.23\%). The streaming model is still more accurate than the vendor's recognizer
(8.5\%/21.3\%) and than Whisper tiny.en (6.3\%/15.9\%) on a laptop. Its full-context mode (\SFl\%/\SFlo\%) loses
about 0.1--0.3 points to the original, the price of noise-robust fine-tuning on read speech.

Noise augmentation matters more than anything else we tried. Averaged over the 14 conditions of
Table~\ref{tab:robust}, the first streaming model, fine-tuned on clean speech with fixed normalization, scored
\SRvfull\% in full-context mode and \SRvstream\% with $C{=}32$ (both with the language model), worse than the
original model's \LMrobust\%. After adding MUSAN noise, music, babble and simulated reverberation (8{,}000 further steps),
the numbers fall to \SRfull\% and \SRstream\%: the streaming model is \emph{more} robust than the original and, at
$C{=}32$, about as robust as Whisper tiny.en (12.1\%) is accurate.

\begin{table}[t]
\centering\small
\caption{One streaming-trained model on the ESP32-S3 engine (int8, with the language model unless stated), against
the original utterance-mode model. Robust.: mean \WER\ over the 14 conditions of Table~\ref{tab:robust}.
$^\ddagger$Measured on the board, one core, six clips of 3.6--10.6\,s: time from the last sample of the clip to the final text, with the streaming rows fed at the rate
of speech and the utterance rows compute only. The 0.8\,s end-of-speech wait comes on top.}
\label{tab:stream}
\setlength{\tabcolsep}{4pt}
\begin{tabular}{@{}lcccc@{}}
\toprule
Mode & clean & other & robust. & final text$^\ddagger$ \\
\midrule
Utterance mode (original) & \LMclean & \LMother & \LMrobust & \BUttOrigLo--\BUttOrigHi\,s \\
Streaming, $C{=}32$, greedy & \SSg & \SSgo & -- & -- \\
Streaming, $C{=}32$ & \SSl & \SSlo & \SRstream & \BPacEnLo--\BPacEnHi\,s \\
Streaming, $C{=}16$ & 6.14 & 13.23 & -- & -- \\
Same model, full context & \SFl & \SFlo & \SRfull & \BUttEnLo--\BUttEnHi\,s \\
\bottomrule
\end{tabular}
\end{table}

\paragraph{Latency.} Two numbers matter. When the chip has kept up with the audio, only the last partial chunk is left, and the final
step takes \BFinSt\,s on average on the board (\BFinStLo--\BFinStHi\,s over 12 clips); in the emulator its instruction count is well described by
$17\,\mathrm{M}+5.5\,\mathrm{M}\times$ (remaining encoder frames), summed over both cores. But the chip does not keep up: at \BRtfStS$\times$ real
time the backlog grows with the utterance. We therefore also fed six clips at the rate of speech, one 20\,ms block when its last sample would have been
spoken, and measured when the final text appears after the last sample: \BPacEnLo--\BPacEnHi\,s (mean \BPacEn\,s) for clips of 3.6--10.6\,s,
against \BUttEnLo--\BUttEnHi\,s for the same clips in utterance mode (the 0.8\,s end-of-speech wait comes on top of both). The delay grows with
the length of the utterance, roughly as $(\mathrm{RTF}-1)\,T$ plus the final step. Streaming roughly halves the wait but does not give the
1.1--1.4\,s we had estimated from instruction counts; that needs a real-time factor below one.

\subsection{Spanish}
\label{sec:spanish}
The engine is language-agnostic: a new language needs a new 1024-piece vocabulary, a re-initialized CTC head and
fine-tuning of the whole network from the English weights. We trained a Spanish model on 2{,}492\,h of Common Voice,
VoxPopuli, MLS and FLEURS~\citep{conneau2023fleurs} (all CC0 or CC-BY), including 1{,}029\,h of Common Voice's unvalidated
clips kept only where NVIDIA's large Spanish Conformer~\citep{nvidia2022ctcsmall} agrees with the reference within 15\%
\WER, with the noise and reverberation augmentation of Section~\ref{sec:streaming}, and a Spanish GRU language model
trained on public-domain books, Common Voice sentences and VoxPopuli transcripts with all development and test sentences
removed. The model needs no engine change; the language model stores its own beam-search weights.

Table~\ref{tab:spanish} compares it with multilingual Whisper tiny on the same utterances (400 evenly spaced per set, 333
for FLEURS after dropping references with digits; same normalization, numerals in Whisper's output spelled out). With the
language model it makes 34--60\% fewer errors on all four sets, and streaming at $C{=}32$ costs about 2 points. The
comparison favors our model on these domains, since we trained on the training splits of the corpora and Whisper is
zero-shot, and we have no Spanish noise benchmark. On the complete test sets (60\,h) the numbers are
13.75/10.90/15.68/11.32\% with the language model. The language-model weights were chosen from a small grid on subsets of
the test sets, on a flat optimum (all settings within 0.2 points of each other).

\begin{table}[t]
\centering\small
\caption{Spanish \WER\ (\%) on 400-utterance subsets (Common Voice 17, MLS, VoxPopuli, FLEURS es\_419).}
\label{tab:spanish}
\setlength{\tabcolsep}{4pt}
\begin{tabular}{@{}lcccc@{}}
\toprule
System & CV & MLS & Vox & FLEURS \\
\midrule
O\'ido + LM (host build) & \textbf{13.3} & \textbf{10.6} & \textbf{15.5} & \textbf{11.3} \\
O\'ido greedy & 20.3 & 14.2 & 20.0 & 16.9 \\
O\'ido streaming $C{=}32$ + LM & 15.2 & 12.1 & 16.3 & 12.8 \\
Whisper tiny (laptop) & 33.0 & 21.5 & 28.7 & 17.1 \\
\bottomrule
\end{tabular}
\end{table}

\subsection{Oído v2: accuracy in streaming mode}
\label{sec:v2acc}
Table~\ref{tab:v2acc} gives the word error rate of the version~2 models on the full LibriSpeech test sets, measured with the host build of the engine, which equals the
chip's arithmetic (18 of 18 board transcripts identical). ``Chunk'' is the streaming chunk length; ``LM 4/6'' is beam 4 with 6 candidates per frame and ``LM 2/3'' the
real-time setting used for the speed measurements, both with the public-domain language model (weight 0.3, bonus 1.0, tuned on dev-clean and two held-out sets).
\begin{table*}[t]
\centering\small
\caption{Oído v2: \WER\ (\%) on LibriSpeech test-clean/test-other. Version~1 streaming model at the same chunk length (0.64\,s, emulation): 8.39/16.14 greedy, and 4.04/8.95 with full context.}
\label{tab:v2acc}
\setlength{\tabcolsep}{3pt}
\begin{tabular}{@{}lccccccc@{}}
\toprule
Model & full, greedy & 0.64\,s, greedy & 1.28\,s, greedy & 0.64\,s, LM 4/6 & 0.64\,s, LM 2/3 & 1.28\,s, LM 4/6 & full, LM 4/6 \\
\midrule
A, 16 blocks, int8 & 5.18/11.60 & 9.28/19.01 & 7.34/15.80 & 6.66/15.32 & 7.29/16.26 & 5.60/13.17 & 4.27/9.92 \\
B, 12 blocks, int8 & 6.56/15.27 & 11.47/23.29 & 9.07/19.84 & 8.38/19.13 & 9.10/20.14 & 6.87/16.46 & 5.25/13.02 \\
A, 16 blocks, int4 & 5.66/12.79 & 10.01/19.90 & 8.02/17.00 & 7.14/16.37 & 7.81/17.29 & 5.96/14.03 & 4.58/10.71 \\
B, 12 blocks, int4 & 7.25/16.38 & 12.28/24.33 & 9.90/20.85 & 8.95/19.85 & 9.76/20.97 & 7.46/17.31 & 5.83/13.75 \\
\bottomrule
\end{tabular}
\end{table*}
The accuracy cost of the faster model is real. At 0.64\,s chunks (greedy) variant A loses 1.6 points on test-clean and 3.8 on test-other against the version~1 streaming model, and variant B
3.9 and 8.2; both recover most of it with 1.28\,s chunks, which are also cheaper to compute (the weights are re-read half as often) and delay the final text by about 0.2\,s more.
Beyond-LibriSpeech sets and noise conditions are not repeated for version~2.

\subsection{Speed and memory on the board (version 1)}
\label{sec:speed}
Table~\ref{tab:speed} gives the measured real-time factor. With one core, the int8 CTC model with the language model runs at
\BRtfEn$\times$ real time on 27 clips of 3.6--10.6\,s (\BRtfEnLo--\BRtfEnHi), the int4 model at \BRtfIv, the transducer at \BRtfRn\ and
the Spanish model at \BRtfEs. The RTF barely depends on the clip, so compute time is proportional to the length of the audio. Figure~\ref{fig:profile}
shows the executed instructions per second of audio in the emulator, measured on 39.7\,s of test-other speech: 220\,M across both
cores, 46\% of them in matrix products and 15\% in attention. The board spends \BCyc\,M cycles per second of audio, \BCpi\ cycles per
executed instruction; Table~\ref{tab:cpi} shows that every stage costs between 1.3 and 2.4 cycles per instruction except the im2col gather (3.7). The time is thus
spread evenly rather than concentrated in a memory-bound stage, and on one core the cost is set by how many instructions the kernels issue, not by waiting
for weights. Beam search with the language model takes 8.7\% of the cycles. If both cores were perfectly parallel, the 55\% of instructions on the critical
path would give an RTF of about \BRtfProj; this is a projection, and the one build we could time with both cores ran at 0.78 times the one-core time
(Section~\ref{sec:dual}).

\begin{table}[t]
\centering\small
\caption{Measured on the board, one core, 240\,MHz, with the language model unless noted. RTF: compute time divided by audio duration (mean,
and range over clips). ``= host'': clips whose transcript equals the host build's. The firmware's transcripts equal the emulator's on all 27 clips of the
first row. Clips: 214.8\,s of audio in the first row, 59--103\,s in the others.}
\label{tab:speed}
\setlength{\tabcolsep}{3.5pt}
\begin{tabular}{@{}lcccc@{}}
\toprule
Model & clips & RTF & range & = host \\
\midrule
CTC int8 (released) & 27 & 1.97 & 1.91--2.02 & 22/27 \\
CTC int4 & 10 & 1.76 & 1.73--1.83 & 8/10 \\
Transducer int8, no LM & 10 & 1.89 & 1.88--1.91 & 10/10 \\
Stream. model, utterance & 12 & 2.04 & 1.98--2.11 & 12/12 \\
Stream. model, $C{=}32$ & 12 & 1.96 & -- & 10/12 \\
Spanish, utterance & 12 & 2.06 & 1.98--2.13 & 9/12 \\
Spanish, $C{=}32$ & 12 & 1.97 & -- & -- \\
\bottomrule
\end{tabular}
\end{table}

\begin{table}[t]
\centering\small
\caption{Cycles on the board against instructions in the emulator, per second of audio (int8 CTC model, 27 clips on the board, 39.7\,s of speech in the
emulator).}
\label{tab:cpi}
\setlength{\tabcolsep}{3pt}
\begin{tabular}{@{}lrrr@{}}
\toprule
Stage & Minstr/s & Mcyc/s & cyc/instr \\
\midrule
log-mel & 10 & 14 & 1.48 \\
conv0 & 14 & 20 & 1.45 \\
im2col & 6 & 20 & 3.66 \\
matrix products & 103 & 216 & 2.09 \\
layernorm & 7 & 15 & 2.07 \\
activation quant & 15 & 32 & 2.16 \\
SiLU / GLU & 11 & 25 & 2.35 \\
q/k/v int8 & 6 & 13 & 2.04 \\
pos. enc. & 3 & 6 & 2.03 \\
attention & 34 & 44 & 1.30 \\
depthwise conv & 10 & 15 & 1.40 \\
head + beam search with LM & 21 & 40 & 1.90 \\
\bottomrule
\end{tabular}
\end{table}

\begin{figure}[t]
\centering
\begin{tikzpicture}[font=\scriptsize, x=0.052cm, y=0.38cm]
\node[anchor=east] at (0,0) {matrix products};
\fill[black!55] (1,-0.32) rectangle (104.4,0.32);
\node[anchor=west] at (104.4,0) {103\,M/s};
\node[anchor=east] at (0,-1) {attention};
\fill[black!55] (1,-1.32) rectangle (34.8,-0.6799999999999999);
\node[anchor=west] at (34.8,-1) {34\,M/s};
\node[anchor=east] at (0,-2) {activation quant};
\fill[black!55] (1,-2.32) rectangle (15.8,-1.68);
\node[anchor=west] at (15.8,-2) {15\,M/s};
\node[anchor=east] at (0,-3) {conv0};
\fill[black!55] (1,-3.32) rectangle (15.2,-2.68);
\node[anchor=west] at (15.2,-3) {14\,M/s};
\node[anchor=east] at (0,-4) {SiLU / GLU};
\fill[black!55] (1,-4.32) rectangle (11.6,-3.68);
\node[anchor=west] at (11.6,-4) {11\,M/s};
\node[anchor=east] at (0,-5) {depthwise conv};
\fill[black!55] (1,-5.32) rectangle (11.5,-4.68);
\node[anchor=west] at (11.5,-5) {11\,M/s};
\node[anchor=east] at (0,-6) {log-mel};
\fill[black!55] (1,-6.32) rectangle (10.7,-5.68);
\node[anchor=west] at (10.7,-6) {10\,M/s};
\node[anchor=east] at (0,-7) {layernorm};
\fill[black!55] (1,-7.32) rectangle (8.0,-6.68);
\node[anchor=west] at (8.0,-7) {7\,M/s};
\node[anchor=east] at (0,-8) {q/k/v int8};
\fill[black!55] (1,-8.32) rectangle (7.4,-7.68);
\node[anchor=west] at (7.4,-8) {6\,M/s};
\node[anchor=east] at (0,-9) {im2col};
\fill[black!55] (1,-9.32) rectangle (6.6,-8.68);
\node[anchor=west] at (6.6,-9) {6\,M/s};
\node[anchor=east] at (0,-10) {pos. enc.};
\fill[black!55] (1,-10.32) rectangle (4.0,-9.68);
\node[anchor=west] at (4.0,-10) {3\,M/s};
\node[anchor=east] at (0,-11) {head + decode};
\fill[black!55] (1,-11.32) rectangle (2.4,-10.68);
\node[anchor=west] at (2.4,-11) {1\,M/s};
\end{tikzpicture}
\caption{Executed instructions per second of audio, by stage, for the int8 CTC model (both cores, exact counts from
instruction-counting emulation).}
\label{fig:profile}
\end{figure}

Table~\ref{tab:ablation} lists the engine optimizations and their effect on executed instructions. The bandwidth-aware
tiling deliberately increases instructions, by 2.9\%, to cut weight traffic 2.4$\times$. Its benefit is invisible to an
instruction-counting emulator, and we did not measure it on silicon (that needs a build without the tiling).

\begin{table}[t]
\centering\small
\caption{Engine optimizations: executed instructions for 39.7\,s of audio (int8 CTC, greedy), cumulative.
$^\dagger$Adds instructions but cuts weight traffic from flash from 18 to 7.5\,MB/s.}
\label{tab:ablation}
\begin{tabular}{@{}lrr@{}}
\toprule
Step & G instr. & $\Delta$ \\
\midrule
Baseline int8 engine & 9.82 & -- \\
+ merged attention score pass & 9.63 & $-$1.9\% \\
+ x-stationary GEMM for $K{\le}256$ & 9.39 & $-$2.5\% \\
+ K-split, batched front end$^\dagger$ & 9.66 & +2.9\% \\
+ int8 positional table, buffer aliasing & 9.58 & $-$0.8\% \\
+ fused im2col quantization & 9.29 & $-$3.0\% \\
+ softmax lookup table & 8.99 & $-$3.2\% \\
+ two-frame depthwise convolution & 8.95 & $-$0.4\% \\
\midrule
Total & & $-$8.9\% \\
\bottomrule
\end{tabular}
\end{table}

Flash holds 14.0\,MB (int8), 8.3\,MB (int4) or 15.5\,MB (transducer), plus 1.3\,MB for the optional language model.
With int4, a 16\,MB module leaves a 6\,MB partition for application code. At the maximum utterance length the
PSRAM working set is 2.4\,MB (emulator), and on the board at least 125\,KB of internal SRAM stays free for the rest of the system (72\,KB with the streaming caches).

\subsection{Latency}
\label{sec:latency}
Table~\ref{tab:latency} gives the time from the last sample of a short voice command to its text, in each mode. In utterance mode it is the compute time, which scales
with the length of the clip at the measured RTF; the 0.8\,s end-of-speech wait comes on top. The streaming rows come from the run fed at the rate of speech
(Section~\ref{sec:streamres}). At present, a 3.7\,s command takes about 7\,s in utterance mode and 4\,s in streaming mode, and neither is
conversational.

\begin{table}[t]
\centering\small
\caption{Time from the last sample of a 3.6--3.8\,s command to its text, measured on the board (one core). $^\dagger$Computed from the measured RTF
(this model was not run on a clip of that length).}
\label{tab:latency}
\begin{tabular}{@{}lcc@{}}
\toprule
Profile & RTF & text after clip \\
\midrule
CTC int8 + LM, utterance mode & 1.97 & 6.9--7.4\,s \\
CTC int4 + LM, utterance mode & 1.76 & 6.2--6.7\,s \\
Transducer int8, utterance mode & 1.89 & 7.0\,s$^\dagger$ \\
Streaming model, utterance mode & 2.04 & 7.2--7.6\,s \\
Streaming model, streaming $C{=}32$ & 1.96 & 4.1--4.2\,s \\
\bottomrule
\end{tabular}
\end{table}

\section{Discussion and Limitations}
\label{sec:discussion}
\paragraph{Where the accuracy comes from.}
The acoustic models are NVIDIA's, trained on NeMo ASRSET, a composite of public English corpora that NVIDIA's model cards put at about 16{,}000 hours~\citep{nvidia2022ctcsmall}. Our contribution is to run them on a
microcontroller without an accelerator and with almost no loss: per-frame dynamic quantization keeps int8 essentially lossless without
retraining, and whole-utterance, block-wise execution lets a 14\,MB model run from flash. We designed the schedule around memory traffic,
amortizing every weight fetch over 64 frames. The board shows, however, that once the model fits and runs, the remaining cost on one core is
instruction issue: about two cycles for each instruction in every stage, with the matrix kernels spending most of their instructions on loads, accumulator
reads and loop overhead rather than on multiply-accumulates. Faster kernels, and the second core, are where the next factor of two lies.

\subsection{The two-core mode}
\label{sec:dual}
In the emulator the two-core schedule is correct, and its instruction counts suggested a real-time factor of 0.7--0.95. On the board, at 240\,MHz, it gives
wrong transcripts and sometimes crashes. We traced it with a hash of every layer's output on one 3.8\,s clip (each configuration was run three times).
One core is exactly reproducible: all 166 hashes are identical in every run and identical to the emulator's. With two cores the activations are corrupted at
random places, always first in layer 1 or 2 and almost always in the feed-forward block of layer 1, and in roughly one run in eight the firmware crashes with an
\texttt{IllegalInstruction} exception and a corrupted backtrace, which suggests corrupted instruction or data fetches rather than a logic error. Two cores at
160\,MHz are correct (166 of 166 hashes identical, in all three runs), but no faster than one core at 240\,MHz.

Parallelizing every part of the model except the matrix products is correct, and parallelizing only the matrix products fails. Among them, the
$K{=}176$ products and the $K{=}3520$ front-end products are correct on two cores, and the $K{=}704$ products (the second linear layer of each feed-forward
block), which move the most weight data per cycle, corrupt. Neither PSRAM speed (80 to 40\,MHz), data-cache line size (64 to 32\,B), the location of the weights
(flash or PSRAM), FreeRTOS semaphores (a spin-wait barrier behaves the same), interrupts (masked while a job runs), the location or sharing of the activation
tile, instruction fetch (kernel in IRAM, or a separate copy per core), nor the split itself (by rows instead of by output channels) removes the errors.
We do not know the cause. A hardware margin problem at 240\,MHz (supply droop or timing) fits the 160\,MHz result, the random values and the crashes,
but an unrelated engine of ours (a text-to-speech engine, not described here) running a synthetic two-core stress of the same shape on the same board at 240\,MHz (K${=}704$, 64 rows,
a shared 45\,KB tile in internal SRAM, an 88/88 split, $3\times90$\,s, plus other shapes, about 21 minutes in all, every call checked against a one-core reference) saw no mismatch and no crash, and its
real firmware passed a 600-iteration soak. A plain supply problem on this board therefore looks unlikely; something specific to our kernel, to how the two
cores are started and joined, or to streaming weights from external memory in lock-step is more likely, and we have not found it. Even a correct two-core mode would not reach real time on this evidence: the incorrect
spin-wait build ran at 0.78 times the one-core time, not the 0.55 that the instruction counts promised, because the cores contend for memory.

\subsection{Limitations}
\begin{itemize}
\item \textbf{Real time costs accuracy.} At 0.64\,s chunks (greedy) variant A is 1.6/3.8 points worse than the version~1 streaming model on test-clean/test-other, and variant B 3.9/8.2;
test-other stays above 10\% in streaming mode for every model we trained (and for version~1). With full context and the language model, variant A reaches 10.7\% on test-other.
\item \textbf{One board, one core.} All speed numbers are from one development board; it dropped off the USB bus three times under sustained load, which we attribute, without
having measured it, to its supply (Section~\ref{sec:dual}). The two-core mode stays disabled.
\item \textbf{The language model is large for this budget.} Its 1.25\,MB are read for every frame that extends a hypothesis; the decoder is the largest single cost after the matrix products.
\item \textbf{Streaming costs accuracy.} Chunked processing loses 4--5 points on test-clean against full context for the version~2 models with greedy decoding and 2.6--3.1 with the language model (Table~\ref{tab:v2acc}).
\item \textbf{Small board study.} We measured one board. The RTF figures cover 27 clips (CTC int8) and 10--12 clips for the other models, all of at most 10.6\,s; the 2.4\,MB
PSRAM peak for a 20\,s utterance comes from the emulator. Accuracy numbers come from the host build; the board's transcripts equal the emulator's, and the host
differs from them on some hard clips (22 of 27 identical), most visibly for Spanish (the board scores 9.3\% against 7.3\% for the host build on 12 clips, 205 words).
\item \textbf{Two languages.} English and Spanish; other languages need a new vocabulary and fine-tuning.
\item \textbf{Transducer license.} The transducer weights are distributed under NVIDIA's NGC terms and are not
redistributed with our release.
\item \textbf{Comparisons.} The laptop baselines use 500-utterance subsets, and MultiNet7 numbers are vendor-reported.
\item \textbf{Framework comparison.} We did not compare against TensorFlow Lite Micro or ESP-DL deployments of the same
model.
\end{itemize}

\section{Conclusion}
O\'ido runs open-vocabulary English speech recognition entirely on a \$5 microcontroller with no neural accelerator. Version~1 was accurate (3.70\%/8.23\% on LibriSpeech with
a 13\,M-parameter Conformer) but ran at twice real time. Measuring the board showed that streaming speech recognition on this chip is bounded by how fast weights can be re-read from
external memory, not by arithmetic. A model built for that memory system (80\,ms frames, a light front end, int4 weights, mostly resident in PSRAM) and an engine that overlaps the weight stream
with compute reach a streaming real-time factor of \VBrtf--\VArtf\ with a language model, with the final text \VBpac--\VApac\,ms after the end of speech, at \VAclean\%/\VAother\% (variant A, 0.64\,s chunks)
and \VAfclean\%/\VAfother\% with full context. That makes a voice command or a conversational turn practical on a single chip; the price is accuracy, most visible on noisy speech. The two-core mode remains
off, and the language model is now one of the largest costs. We release the engine, firmware, models and the board measurements so that others can build on them and check them.
The code is licensed under GPLv3; the models are CC-BY-4.0 (int8 and Spanish) or CC-BY-SA-4.0 (int4 and streaming), as derivatives of NVIDIA's CC-BY-4.0 model.

\section*{Acknowledgments}
We thank NVIDIA for releasing the NeMo Conformer models under open terms, and the creators of the datasets used for
evaluation.
Parts of the engine code, the experiment scripts and this manuscript were drafted with the help of an AI assistant
(Claude, Anthropic). The authors directed the work, checked the results and the text, and take full responsibility for
the content.

{\small
\bibliographystyle{plainnat}
\bibliography{refs}
}

\end{document}